\documentclass{article}

\usepackage[dblblindworkshop, final]{neurips_2025}

\usepackage[utf8]{inputenc} 
\usepackage[T1]{fontenc}    
\usepackage{hyperref}       
\usepackage{url}            
\usepackage{booktabs}       
\usepackage{amsfonts}       
\usepackage{nicefrac}       
\usepackage{microtype}      
\usepackage{xcolor}         
\usepackage[table]{xcolor}
\usepackage{array}
\usepackage{booktabs}
\usepackage{tcolorbox}
\usepackage{amsmath}
\usepackage{mdframed}
\usepackage{enumitem}

\usepackage{algorithm}
\usepackage{algpseudocode}
\usepackage{amssymb}
 
\title{FireGNN: Neuro-Symbolic Graph Neural Networks with Trainable Fuzzy Rules for Interpretable Medical Image Classification}

\author{%
  Prajit Sengupta \\
  BASIRA Lab\\
  Department of Computing \\
  Imperial College London \\
  \texttt{prajit.sengupta06@gmail.com} 
  \And
  Islem Rekik\thanks{Corresponding author: \url{i.rekik@imperial.ac.uk}, BASIRA Lab: \url{https://basira-lab.com} \\ \textbf{Repository}: \url{https://github.com/basiralab/FireGNN}} \\
  BASIRA Lab\\
  Department of Computing \\
  Imperial College London \\
  \texttt{i.rekik@imperial.ac.uk} \\
}

\begin{document}

\maketitle

\begin{abstract}
Medical image classification requires not only high predictive performance but also interpretability to ensure clinical trust and adoption. Graph Neural Networks (GNNs) offer a powerful framework for modeling relational structures within datasets, however, standard GNNs often operate as black boxes, limiting transparency and usability particularly in clinical settings. In this work, we present an interpretable graph-based learning framework named \textbf{FireGNN} that integrates \textit{trainable fuzzy rules into GNNs} for medical image classification. These rules embed topological descriptors - node degree, clustering coefficient, and label agreement - using learnable thresholds and sharpness parameters to enable intrinsic symbolic reasoning. Additionally, we explore \textit{auxiliary self-supervised tasks} (e.g., homophily prediction, similarity entropy) as a benchmark to evaluate the contribution of topological learning. Our fuzzy-rule-enhanced model achieves strong performance across five MedMNIST benchmarks and the synthetic dataset MorphoMNIST, while also generating interpretable rule-based explanations. To our knowledge, this is the first integration of trainable fuzzy rules within a GNN. 
\end{abstract}

\section{Introduction}

Medical image classification is fundamental to clinical decision-making and diagnostic workflows. While deep learning models, particularly Convolutional Neural Networks (CNNs), have demonstrated strong performance on medical imaging benchmarks, their lack of interpretability remains a key barrier to clinical adoption. In high-stakes domains like healthcare, it is crucial for models to not only make accurate predictions but also provide transparent, human-understandable reasoning behind those decisions \cite{graphAI_medicine,SALAHUDDIN2022105111}. Graph Neural Networks (GNNs) have emerged as a compelling alternative by capturing topological and relational patterns across medical datasets \cite{howpowerfulgnn}. By representing each image as a node and connecting similar samples via edges (e.g., using cosine similarity in feature space), GNNs can model global structure often overlooked by CNNs \cite{graph_meddiag,brain_gnn_interpret}. However, standard GNN architectures such as Graph Convolutional Networks (GCNs) face major limitations \cite{pitfalls_gnn}. 
\textbf{First}, GNNs often operate as black boxes, offering little insight into how predictions are made. For example, when a tumor patch is classified as malignant, it is unclear what node features or connections most influenced that decision - posing a major trust issue in clinical settings \cite{Castellano}. \textbf{Second}, many GNN pipelines use simplistic heuristics (e.g., spatial proximity) to construct graphs, especially in histopathology \cite{GNNhistopathology,design_space_gnn}. These hand-crafted rules may ignore biologically meaningful structures, leading to suboptimal or biased learning. \textbf{Third}, GNNs typically assume fixed graph topologies \cite{heterophiliy_topological}, making it difficult to adapt to new data without full retraining. This rigidity limits scalability in evolving clinical environments \cite{graphAI_medicine,medimg_mdpi}. These limitations motivate our proposed solution: \textbf{FireGNN} - a \textbf{Fuzzy Interpretable Rule Embedding for GNNs}. Unlike prior interpretability methods that work post-hoc, \textbf{FireGNN} integrates symbolic reasoning directly into the forward pass of a GNN. It does this by embedding interpretable fuzzy rules over key \textbf{node-level topological features} - specifically, the \textit{node's degree}, \textit{clustering coefficient}, and \textit{2-hop label agreement}.
Each fuzzy rule in FireGNN is parameterized by a trainable threshold and sharpness factor, enabling the model to define concepts like \textit{“high connectivity”} or \textit{“strong label consistency”} in a data-driven learnable way. These rules then \textit{"fire”} with a certain strength, and are fused with learned graph embeddings through a gating mechanism. The result is a \textbf{transparent and expressive model} that can say, for example:
\vspace{-2mm}
\begin{quote}
\textit{``This node is classified as a liver region due to its high degree and strong 2-hop label agreement with neighboring nodes.''}
\end{quote}
\vspace{-2mm}

\noindent As an exploratory benchmark, we also evaluate whether \textbf{auxiliary self-supervised tasks}-such as prediction of local homophily and similarity entropy—can guide GNNs toward more topology-aware embeddings. However, we find that these tasks yield only modest gains compared to the symbolic reasoning introduced by fuzzy rules. FireGNN is evaluated on five MedMNIST datasets and one synthetic dataset (MorphoMNIST), where it achieves state-of-the-art accuracy while offering intrinsic interpretability. The key contributions of our paper can be summarized as follows:
\begin{enumerate}
    \item \textit{On a methodological level:} We present FireGNN that integrates trainable fuzzy rules over topological descriptors, enabling data-adaptive symbolic reasoning fused with learned embeddings for classification.
    \item \textit{On a clinical level:} Our intrinsically interpretable framework supports trustworthy medical AI, offering human-readable explanations (e.g., “high label agreement → liver node”) alongside high performance, encouraging real-world deployment.
    \item \textit{On a generic level:} Our model generalizes across various datasets and GNN backbones \textit{(GCN, GAT, GIN)}, and can be extended to any graph domain requiring explainable structural modeling. We also benchmarked against auxiliary tasks as an alternative way to encode topological structure but found limited improvement.
\end{enumerate}

\section{Related Work}
\textbf{Fuzzy logic in graph-based learning} Fuzzy logic provides a framework for modeling uncertainty and interpretability using linguistic rules and soft membership functions. Classical systems like ANFIS \cite{FuzzyLogic,ANFIS} combine fuzzy rule-based reasoning with trainable neural architectures. In the context of graphs, FGNN \cite{wei_fuzzy} leverages fuzzy inference for few-shot learning, and FL-GNN \cite{inbook} integrates fuzzy systems into message-passing networks. Prior work such as \cite{Castellano} has explored combining fuzzy logic with GNNs to enhance model transparency. These methods have demonstrated the potential of fuzzy logic for enhancing interpretability, though they often rely on fixed rule templates.

\noindent \textbf{Symbolic reasoning \& structure-aware GNNs}
Incorporating symbolic rules into neural models has been explored as a path toward human-understandable AI \cite{gnn_neurosymbolic}. However, many fuzzy and symbolic GNNs use pre-defined thresholds or rules that do not adapt to different datasets \cite{fuzzy_rules,extract_fuzzy}. Parallel to this, several studies have investigated topology-aware learning using auxiliary tasks or edge-type decoupling. \cite{manessi} proposed using structural signals like homophily for improved node classification, while DuoGNN \cite{DuoGNN} decouples edges into homophilic vs. heterophilic sets and processes each through parallel GNN streams. Our model builds on this by embedding fuzzy logic over graphs.

\noindent \textbf{Intrinsic vs post-hoc interpretability in GNNs}
Post-hoc explanation methods such as GNNExplainer \cite{GNNExplainer} and PGExplainer aim to highlight important substructures or features after training. While useful, these approaches operate outside the model and may not faithfully reflect its internal reasoning \cite{Liu2022}. Intrinsically interpretable models, by contrast, aim to embed explanation directly within the forward computation, offering more transparent and stable reasoning pipelines—especially critical in medical and safety-critical domains. Throughout this paper we are formulating \textbf{2 hypotheses}:

\noindent \textbf{H1: Rule-based interpretability} \textit{Embedding fuzzy rules over node-level structural features—such as \textit{degree}, \textit{clustering coefficient}, and \textit{label agreement}—enables the model to generate intrinsic, human-readable explanations.}

\noindent \textbf{H2: Trainable symbolic reasoning} \textit{We hypothesize that learning fuzzy rule thresholds ($\theta$) and sharpness parameters ($\alpha$) from data allows the model to adapt boundaries to the specific graph structure, improving both generalization and semantic alignment.}

\section{Methodology}
\noindent Existing GNNs either ignore rich topological signals or treat them as fixed inputs, and prior fuzzy-GNN hybrids rely on hand-crafted thresholds.

\begin{mdframed}[frametitle={\colorbox{white}{\quad Key challenge:\quad}},
frametitleaboveskip=-\ht\strutbox,
frametitlealignment=\center
]
\vspace{-2mm}
 How can we learn \emph{data-adaptive} symbolic conditions over graph-theoretic features (e.g., degree, clustering, etc) that both improve classification accuracy and yield \emph{intrinsic interpretability}?
\end{mdframed}

\begin{figure}[t]
    \centering
    \includegraphics[width=1\linewidth]{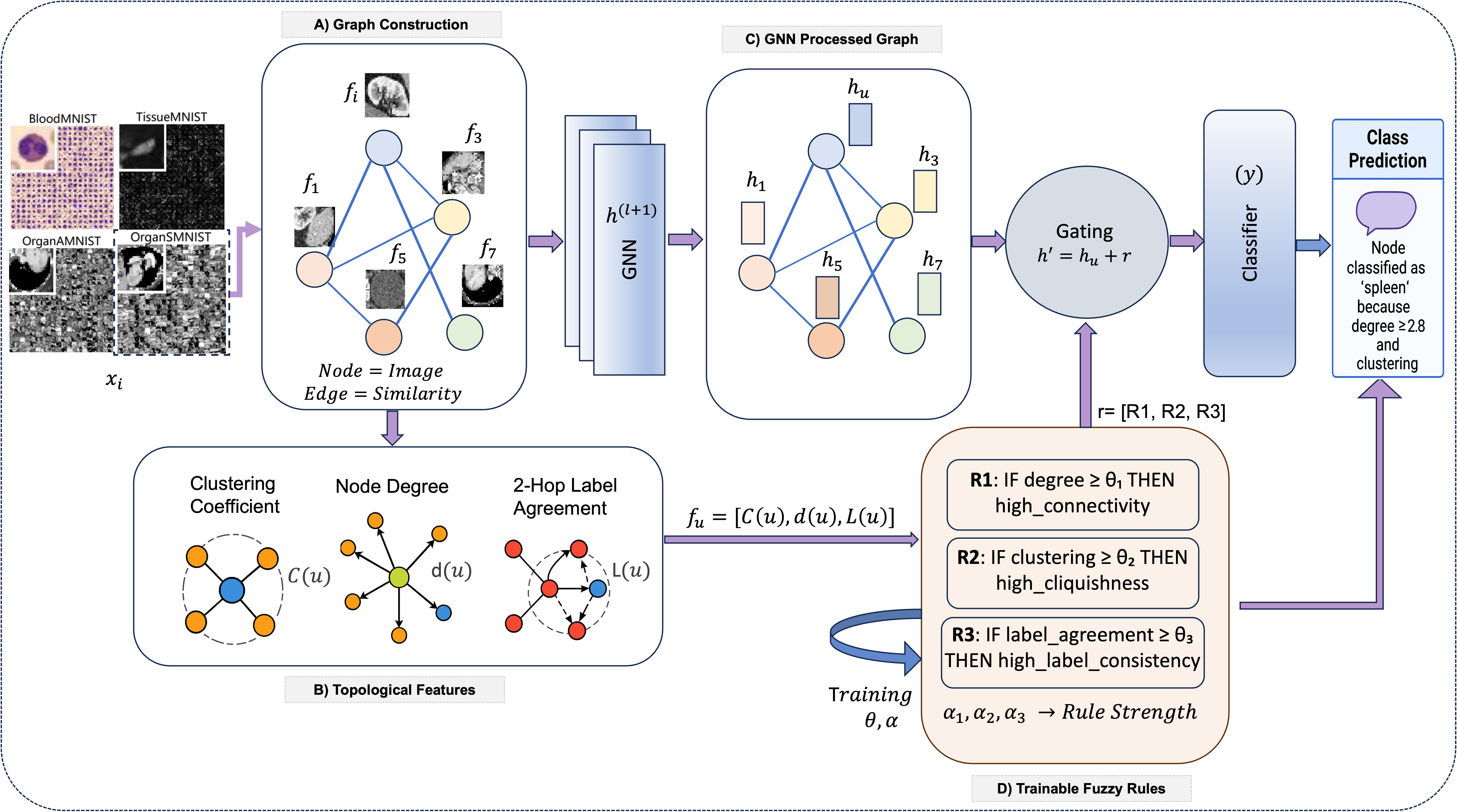}
    \caption{FireGNN architecture with four main components: (A) Graph construction, (B) Extraction of topological features, (C) GNN processing, and (D) Trainable fuzzy rules.}
    \label{fig:FireGNN_flow}
\end{figure}

\noindent To this end, we propose \textbf{FireGNN}, a principled framework that (i) discovers, during training, the precise thresholds and sharpness for each rule per dataset, and (ii) integrates structural reasoning directly into the GNN’s forward pass using symbolic rules over node degree, clustering coefficient, and 2-hop label agreement. An overview of the architecture as shown in \textbf{Fig.~\ref{fig:FireGNN_flow}} is described below:

\noindent \textbf{A. Graph construction} We form a graph $G = (V, E)$ over all images: each node $v_i$ has feature $f_i = F(x_i)$ and label $y_i$ (Fig.~\ref{fig:FireGNN_flow}-A). The adjacency matrix $A$ is built via top-$k$ cosine similarity. The base GCN processes the graph in the usual way as in (Fig.~\ref{fig:FireGNN_flow}-C). Let $H^{(0)} = X$ be the initial node feature matrix. At each layer $\ell$, we compute:
\begin{equation}
H^{(\ell + 1)} = \sigma\left( \tilde{D}^{-1/2} \tilde{A} \tilde{D}^{-1/2} H^{(\ell)} W^{(\ell)} \right)  
\end{equation}
\noindent where $\tilde{A} = A + I$ (adjacency with self-loops), $\tilde{D}$ is the degree matrix of $\tilde{A}$, and $W^{(\ell)}$ are trainable weight matrices. After $L$ layers, we obtain the final node embeddings $h_v$. 


\noindent \textbf{B. GNN with trainable fuzzy rules} Our first model as shown in Fig.~\ref{fig:FireGNN_flow} augments the GNN with a neuro-symbolic fuzzy rule module. For each node $u$, we form a fact vector $f_u \in \mathbb{R}^3$ as in (Fig.~\ref{fig:FireGNN_flow}-B) containing its topological features:
\begin{equation}
f_u = [d(u),\, C(u),\, L(u)],
\end{equation}
\noindent where $d(u)$ is the node degree, $C(u)$ is the clustering coefficient (the ratio of existing edges among $v$’s neighbors to the maximum possible), and $L(u)$ is the 2-hop label agreement. We define three fuzzy rules corresponding to these features (Fig.~\ref{fig:FireGNN_flow}-D). Each rule $i$ has a learnable threshold $\theta_i$ and sharpness $\alpha_i$, where $\theta_i$ controls the decision boundary and $\alpha_i$ modulates how sharply the rule activates near that threshold (i.e, how crisp or gradual the transition is). The activation of rule $i$ on node $u$ is computed:
\begin{equation}
r_i(u) = \sigma\left( \alpha_i (f_u[i] - \theta_i) \right) \in [0,1] 
\end{equation}
\noindent where $\sigma$ denotes the sigmoid function. Thus, $r_i(u)$ is close to 1 when the feature $f_u[i]$ significantly exceeds its threshold. Collecting all rule activations yields the firing strength vector: \(
r(u) = [r_1(u),\, r_2(u),\, r_3(u)] \in [0,1]^3.
\)
\noindent These activations serve as a symbolic explanation of the node’s structural role. For instance, $r_1(u) \approx 1$ suggests ``node has high degree''. We fuse the fuzzy outputs with the GNN embedding  via a gating mechanism. First, project the rule vector into the GNN embedding space:
\begin{equation}
e_u = W_r\, r(u) + b_r, \quad e_u \in \mathbb{R}^d 
\end{equation}
\noindent Then, concatenate $[h_u \parallel e_u] \in \mathbb{R}^{2d}$ and compute a gate vector:
\begin{equation}
g_u = \sigma(W_g [h_u \parallel e_u] + b_g) \in [0,1]^d
\end{equation}
\noindent The updated node embedding is given by:
\begin{equation}
h'_u = g_u \odot h_u + (1 - g_u) \odot e_u  
\end{equation}
\noindent where $\odot$ denotes elementwise multiplication. Intuitively, if rule-based information is highly relevant, $g_u$ increases the influence of $e_u$; otherwise, it favors the original embedding $h_u$. The final embedding $h'_u$ is passed to a linear classifier for label prediction. The entire model, including GCN weights, fuzzy rule parameters $\{\theta_i, \alpha_i\}$, and fusion weights $(W_r, W_g)$, is trained end-to-end by minimizing the cross-entropy loss on labeled nodes. These rules offer interpretability: one can inspect which rules fired to explain a node’s label prediction (e.g.,“high\_connectivity” for bladder class).




\noindent \textbf{C. GNN with auxiliary tasks as benchmark} To assess alternative ways of injecting structural information into the GNN, we also experiment with two auxiliary prediction tasks which serves solely as a \textbf{benchmarking baseline} for comparison. Specifically, we supervise the model to estimate (1) \textit{homophily} \( h(v) \)-the proportion of neighbors sharing the same label as node \( v \), and (2) \textit{similarity entropy} \( S(v) \)-a measure of uncertainty in edge-weight distributions over \( v \)'s neighborhood. These quantities capture key graph properties: label agreement and connection sharpness, respectively. Detailed formulations, and the auxiliary-task training procedure 
are provided in \textbf{Appendix A.2}.

\section{Results and Discussion}
We conduct a comprehensive evaluation of our models on six inductive graph datasets as described in Table~\ref{datsets} (\textbf{Appendix A.1}): five derived from MedMNIST (OrganCMNIST, OrganAMNIST, OrganSMNIST, TissueMNIST, BloodMNIST) \cite{medmnist_data} and one synthetic benchmark (Morpho-MNIST) \cite{morpho_mnist}. To ensure robust assessment, we perform 3-fold cross-validation with three random seeds (42,43,44), yielding a total of nine experiments per model variant. All results report mean±standard deviation across these runs. The complete end-to-end training procedure is as Algorithm 1 (\textbf{Appendix A.8}). All hyperparameters for the GNNs, fuzzy rule module, are detailed in \textbf{Appendix A.4} (Tables 4-7).

To enhance interpretability and transparency, we provide animated visualizations illustrating how fuzzy rule values evolve across epochs during training, along with a demo video showing the convergence of the learnable $\theta$ parameters for each dataset. These materials complement the reported quantitative results by offering an intuitive view of the model’s self-reasoning and rule adaptation dynamics. Additionally, the repository includes all processed datasets and splits used in our experiments, enabling full reproducibility and ease of benchmarking. All supplementary videos, datasets, and resources are available at our project repository: \url{https://github.com/basiralab/FireGNN}.


\noindent \textbf{Learned fuzzy rules} 
Our model discovers dataset-driven fuzzy rule thresholds through training. On OrganCMNIST, the learned means (across 9 runs) are:
\[
\theta_1 = 7.28,\quad \theta_2 = 0.18,\quad \theta_3 = 0.67.
\]
\definecolor{lightpink}{RGB}{255, 230, 235}

\begin{mdframed}[frametitle={\colorbox{lightpink}{\quad Learned Fuzzy Rules (OrganCMNIST)\quad}},frametitlealignment=\center]
\begin{itemize}[leftmargin=*,noitemsep]
  \item \textbf{Rule 1:} IF degree $\ge 7.28$ THEN high\_connectivity  
  \item \textbf{Rule 2:} IF clustering $\ge 0.18$ THEN high\_cliquishness  
  \item \textbf{Rule 3:} IF 2-hop label agreement $\ge 0.67$ THEN high\_label\_consistency  
\end{itemize}
\end{mdframed}
\noindent For example, a node predicted as \emph{bladder} had a degree of 10, a clustering coefficient of 0.18, and a 2-hop label agreement of 0.73. Since:
\begin{itemize}[noitemsep, leftmargin=*]
    \item degree = 10 $\ge$ $\theta_1 = 7.28$ $\Rightarrow$ Rule 1 activated (strength = 0.60),
    \item clustering = 0.18 = $\theta_2$ $\Rightarrow$ Rule 2 marginally activated (strength = 0.50),
    \item label agreement = 0.73 $\ge$ $\theta_3 = 0.67$ $\Rightarrow$ Rule 3 strongly activated (strength = 0.56),
\end{itemize}
interprets this node as having high connectivity and strong label consistency, providing a human-readable justification for its \emph{bladder} classification. This supports \textbf{H1} by offering intrinsic interpretability through explicit rule activations and \textbf{H2} by adapting thresholds to dataset-driven structural patterns.

\begin{table}[h]
\centering
\scriptsize  
\setlength{\tabcolsep}{2.0pt}
\renewcommand{\arraystretch}{1.2}
\caption{Comparison of GNN variants on six inductive datasets.}
\label{tab:gcnsixdatasets}

  \scalebox{1.0}{\begin{tabular}{l
  *{2}{>{\centering\arraybackslash}p{1.3cm}
        >{\centering\arraybackslash}p{1.3cm}
        >{\centering\arraybackslash}p{1.3cm}
        >{\centering\arraybackslash}p{1.3cm}}}
\toprule
& \multicolumn{4}{c}{\bfseries OrganCMNIST}
& \multicolumn{4}{c}{\bfseries OrganAMNIST} \\
\cmidrule(lr){2-5}\cmidrule(lr){6-9}
\bfseries Method
  & ACC & F1   & Sensitivity  & ROC-AUC
  & ACC & F1   & Sensitivity  & ROC-AUC \\
\midrule
GCN & 88.20±0.61 & 86.03±0.89 & 86.13±0.06 & 99.09±0.08
    & 91.85±0.30 & 91.19±0.33 & 91.18±0.03 & 99.51±0.03 \\
GCN+ Aux & 88.41±0.44 & 86.48±0.58 & 86.43±0.04 & 98.84±0.10 & 93.11±0.24 & 92.60±0.26 & 92.50±0.02 & 99.28±0.05 \\
\rowcolor{green!40}
GCN+ FR & 91.41±0.61 & 89.71±0.58 & 89.74±0.06 & 99.54±0.05 & 94.32±0.18 & 93.88±0.18 & 93.81±0.02 & 99.77±0.01 \\
\arrayrulecolor{gray}\hline
GAT & 90.31±0.28 & 88.47±0.34 & 88.51±0.03 & 99.38±0.05 & 93.69±0.36 & 93.26±0.28 & 93.36±0.04 & 99.69±0.02\\
GAT+ Aux & 90.88±0.49 & 89.07±0.59 & 88.93±0.62 & 99.45±0.08 & 93.70±0.46 & 93.36±0.44 & 93.33±0.04 & 99.54±0.02\\
\rowcolor{green!40}
GAT+ FR & 91.66±0.48 & 90.02±0.52 & 90.12±0.05 & 99.56±0.05 &  94.52±0.31 & 94.08±0.30 & 94.13±0.03 & 99.78±0.02 \\
\arrayrulecolor{gray}\hline
GIN & 87.96±0.59 & 85.61±0.75 & 85.56±0.75 & 98.86±0.09 & 91.54±0.71 & 90.45±0.73 & 90.50±0.69 & 99.41±0.08 \\
GIN+ Aux & 88.53±0.44 & 86.37±0.48 & 86.37±0.48 & 98.65±0.13 & 92.18±0.33 & 91.16±0.39& 91.13±0.39 & 99.26±0.05 \\
\rowcolor{green!40}
GIN+ FR &  89.12±1.18 & 86.81±1.51 & 86.82±1.56 & 99.38±0.16 & 92.48±1.82 & 91.32±2.69 & 91.27±2.71 & 99.59±0.33 \\
\bottomrule
\end{tabular} }

\vspace{0.2em}

  \scalebox{1.0}{\begin{tabular}{l
  *{2}{>{\centering\arraybackslash}p{1.3cm}
        >{\centering\arraybackslash}p{1.3cm}
        >{\centering\arraybackslash}p{1.3cm}
        >{\centering\arraybackslash}p{1.3cm}}}
\toprule
& \multicolumn{4}{c}{\bfseries OrganSMNIST}
& \multicolumn{4}{c}{\bfseries TissueMNIST} \\
\cmidrule(lr){2-5}\cmidrule(lr){6-9}
\bfseries Method
  & ACC & F1   & Sensitivity  & ROC-AUC
  & ACC & F1   & Sensitivity  & ROC-AUC \\
\midrule
GCN & 78.62±0.82 & 73.74±0.99 & 73.85±0.08 & 97.80±0.11
    & 50.90±0.32 & 32.61±0.79 & 32.51±0.07 & 81.98±0.31 \\
GCN+ Aux & 79.19±0.74 & 74.21±0.84 & 74.28±0.07 & 97.56±0.14 & 52.70±0.22 & 35.67±0.36 & 35.60±0.04 & 82.99±0.19 \\
\rowcolor{green!40}
GCN+ FR & 85.05±0.43 & 80.56±0.61 & 80.74±0.04 & 98.95±0.05 &  65.73±0.88 & 51.63±1.22 & 51.59±0.09 & 93.56±0.36 \\
\arrayrulecolor{gray}\hline
GAT & 81.80±0.68 & 77.22±0.73 & 77.16±0.07 & 98.39±0.09 & 51.53±0.35 & 33.10±0.56 & 33.06±0.07 & 83.11±0.12 \\
GAT+Aux & 81.69±0.68 & 77.33±0.73 & 77.06±0.07 & 98.28±0.09 & OOM & OOM & OOM & OOM \\
\rowcolor{green!40}
GAT+FR & 84.82±0.52 & 80.23±0.80 &  80.43±0.05 & 98.93±0.07 & \cellcolor{white}OOM & \cellcolor{white}OOM & \cellcolor{white}OOM & \cellcolor{white}OOM \\
\arrayrulecolor{gray}\hline
GIN &  77.23±0.62 & 71.65±0.65 & 71.77±0.76 & 97.36±0.10 & 50.51±1.09 & 30.31±3.70 & 32.50±2.05 & 81.72±0.85 \\
GIN+Aux &  79.26±1.02 & 73.73±1.97 & 74.29±1.23 & 97.44±0.11 & 51.31±1.42 & 31.12±3.54 & 33.20±1.70 & 82.42±0.27\\
\rowcolor{green!40}
GIN+ FR & 81.46±3.14 & 76.69±3.85 & 76.81±3.84 & 98.51±0.52 & 64.07±2.44 & 48.59±3.93 & 48.59±3.96 & 92.84±1.25 \\
\bottomrule
\end{tabular}}

\vspace{0.2em}

  \scalebox{1.0}{\begin{tabular}{l
  *{2}{>{\centering\arraybackslash}p{1.3cm}
        >{\centering\arraybackslash}p{1.3cm}
        >{\centering\arraybackslash}p{1.3cm}
        >{\centering\arraybackslash}p{1.3cm}}}
\toprule
& \multicolumn{4}{c}{\bfseries BloodMNIST}
& \multicolumn{4}{c}{\bfseries MorphoMNIST} \\
\cmidrule(lr){2-5}\cmidrule(lr){6-9}
\bfseries Method
  & ACC & F1   & Sensitivity  & ROC-AUC
  & ACC & F1   & Sensitivity  & ROC-AUC \\
\midrule
GCN & 80.49±0.66 & 77.46±0.96 & 77.15±0.09 & 96.56±0.18
    & 90.89±0.05 & 90.82±0.02 & 90.73±0.02 & 98.88±0.03 \\

GCN+ Aux & 80.17±0.66 & 77.07±0.84 & 76.96±0.10 & 96.05±0.13 & 92.84±0.10& 92.83±0.10 &  92.78±0.03 & 99.00±0.04 \\
\rowcolor{green!40}
GCN+ FR & 88.31±0.37 & 86.36±0.45 & 86.22±0.04 & 99.15±0.05 & 94.76±1.02 & 94.76±1.03 & 94.58±0.34 & 99.63±0.25 \\
\arrayrulecolor{gray}\hline
GAT & 82.02±0.31 & 79.38±0.40 & 79.18±0.04 & 97.45±0.08 & 91.50±0.15 & 91.48±0.16 & 91.33±0.05 & 98.73±0.04 \\
GAT+ Aux & 81.87±2.08 & 79.70±2.21 & 79.73±0.23 & 97.13±0.46 & OOM & OOM & OOM & OOM \\
\rowcolor{green!40}
GAT+ FR & 87.79±2.08 & 85.86±2.21 & 85.80±0.23 & 99.01±0.46 & \cellcolor{white}OOM & \cellcolor{white}OOM & \cellcolor{white}OOM & \cellcolor{white}OOM \\
\arrayrulecolor{gray}\hline
GIN &  80.30±0.60 & 77.21±0.91 & 76.44±0.95 & 96.58±0.21 & 91.60±0.19 & 91.59±0.18 & 91.60±0.19 & 98.75±0.05 \\
GIN+ Aux &  80.47±0.20 & 77.13±1.31 & 76.83±0.38  & 96.30±0.42 & 92.30±0.24 & 92.11±0.32 & 92.50±0.43 & 99.25±0.12 \\
\rowcolor{green!40}
GIN+ FR & 84.83±3.04 & 82.41±3.60 & 82.39±3.81 & 98.36±0.94 & 93.72±1.55 & 93.73±1.54 & 93.72±1.55 & 99.55±0.22 \\
\bottomrule
\end{tabular}}
\begin{minipage}{2\columnwidth}
\vspace{0.1cm}
\vspace{0.1cm}
\small\raggedright Notes: \textbf{Aux} refers to Auxiliary Tasks; \textbf{FR} denotes Fuzzy Rules; Best Results in \textbf{Green}.
\end{minipage}

\end{table} 
\noindent \textbf{Classification performance}
Table~\ref{tab:gcnsixdatasets} details accuracy (ACC), macro F\textsubscript{1}, sensitivity, and ROC-AUC for each model. On OrganCMNIST, GCN+Fuzzy attains \textbf{91.41\% ±0.61} ACC (+3.21\% over GCN) and 89.71±0.58 F\textsubscript{1}. GCN+Aux achieves a modest accuracy of 88.41±0.44, which even \textbf{sometimes decreases or remains unchanged}, indicating that \textit{auxiliary tasks alone yield only incremental improvements}. This supports our hypothesis that symbolic rule-based modeling is a more effective approach. GAT+Fuzzy further increases OrganCMNIST ACC to 91.66±0.48, confirming the generality of fuzzy reasoning across backbones.  On low-homophily TissueMNIST, GCN+Fuzzy improves ACC by 14.83\% (65.73±0.88 vs 50.90 ±0.32), underscoring its robustness where standard GCNs falter. Similarly, GCN+Aux enhances performance (52.50±0.26 ACC). We also tested the fuzzy rule methodology on Morpho-MNIST, a synthetic dataset, designed to evaluate representation learning. Morpho-MNIST results similarly favor GCN+ Fuzzy (94.76 ± 1.02 ACC), demonstrating generalizability. The \textbf{computational efficiency \& dynamics} are described in Table~\ref{tab:efficiency}(\textbf{Appendix A3}). 

\noindent \textbf{Conclusion} We presented \textbf{FireGNN}, a simple yet effective framework that integrates trainable fuzzy rules for interpretable and accurate classification (Table \ref{tab:gcnsixdatasets}). We also evaluated auxiliary structural tasks as a benchmarking tool for topology encoding. While our fuzzy rules capture key topological cues, future extensions could explore learning richer or more complex rule types. Expanding beyond the three structural features we use and applying this approach to dynamic or heterogeneous graphs offers promising directions for broader and more flexible interpretability \textbf{[see Appendix A.6 (Table 9) for learned rule thresholds, A.7 for a detailed exploration of other theoretical rules]}. Additionally, improving the efficiency of FireGNN particularly reducing its average epoch time and peak memory usage than the current values reported in Table \ref{tab:efficiency} remains an important step toward sustainable model.


\small
\bibliographystyle{plain}
\bibliography{references}





\newpage
\appendix

\section{Technical Appendices and Supplementary Material}

\subsection{Dataset Overview}
The datasets used in our experiments span both real biomedical image collections \textbf{MedMNIST v2} (MedMNIST v2 is available under the CC BY 4.0 license) which is a large-scale collection of lightweight 2D biomedical image datasets covering diverse organ types and 
a synthetic benchmark \textbf{Morpho-MNIST} (Morpho-MNIST is derived from MNIST, which is available under the CC BY-SA 3.0 license). Each dataset is converted into a graph where images serve as nodes 
and edges are formed via top-$k$ cosine similarity in the feature space. This allows us to model global 
topological relationships between samples, which are often overlooked in purely CNN-based approaches.

\begin{table}[h]
\centering
\caption{Graph dataset details created from images}
\label{datsets}
\scalebox{0.71}{
\begin{tabular}{|l|c|c|c|c|c|c|}
\hline
 & \textbf{OrganCMNIST} & \textbf{OrganAMNIST} & \textbf{OrganSMNIST} & \textbf{TissueMNIST} & \textbf{BloodMNIST} & \textbf{Morpho‑MNIST} \\
\hline
\# of Nodes & 23583 & 58830 & 25211 & 236386 & 17092 & 280000 \\
\# of Edges & 82315 & 205404 & 88951 & 826714 & 60116 & 970699 \\
\# of Features & 512 & 512 & 512 & 512 & 512 & 512 \\
\# of Labels & 11 & 11 & 11 & 8 & 8 & 4 \\
Task Type & Multi-class & Multi-class & Multi-class & Multi-class & Multi-class & Multi-class \\
Training Type & Inductive & Inductive & Inductive & Inductive & Inductive & Inductive \\
\hline
Training Nodes & 12975 & 34561 & 13932 & 165466 & 11959 & 216000 \\
Validation Nodes & 2392 & 6491 & 2452 & 23640 & 1712 & 24000 \\
Test Nodes & 8216 & 17778 & 8827 & 47280 & 3421 & 40000 \\
\hline
\end{tabular}}
\end{table}

As summarized in Table~\ref{datsets}, dataset sizes vary substantially: BloodMNIST contains around 17K nodes, while 
TissueMNIST scales to over 230K nodes with nearly 830K edges. This diversity tests FireGNN’s scalability. 
Morpho-MNIST, while synthetic, is deliberately challenging because its classes differ by morphological 
attributes (e.g., thickness - thick, thin; fragmentation - broken), contains 280K nodes, 970K edges, and 4 classes, requiring the model to rely on structural patterns rather than 
simple pixel statistics.

\subsection{GNN with Auxiliary Tasks Mathematical Formalization}
The purpose of these auxiliary tasks is to explicitly encourage the model to be aware of graph topology 
during training, beyond the main classification objective. The homophily term $h(v)$ captures whether 
a node shares labels with its neighbors, which reflects the level of semantic alignment in the graph. 
The similarity entropy $S(v)$, by contrast, quantifies the sharpness of a node’s neighborhood distribution, 
distinguishing between nodes with a few strong connections versus many diffuse ones. 

\noindent Given a node \( v \), we define homophily as  
\( h(v) = \frac{1}{|N(v)|} \sum_{u \in N(v)} \mathbb{I}(y_u = y_v) \),  
where \( N(v) \) is the set of neighbors of \( v \), and \( y_u \) denotes the label of node \( u \). For similarity entropy, edge weights are computed as  
\[ w_{vu} = \frac{\exp(-\|x_v - x_u\|^2)}{\sum_{k \in N(v)} \exp(-\|x_v - x_k\|^2)}  \]  
followed by entropy computation:  
\[ S(v) = -\sum_{u \in N(v)} w_{vu} \log w_{vu} \]  
Two lightweight MLP heads are added to the GNN to predict \( \hat{h}(v) \) and \( \hat{S}(v) \) from \( h_v \), optimized via mean squared error:  
\begin{equation}
\mathcal{L}_{\text{aux}} = \frac{1}{|\mathcal{V}_L|} \sum_{v \in \mathcal{V}_L} \left( (\hat{h}(v) - h(v))^2 + (\hat{S}(v) - S(v))^2 \right)
\end{equation}

\noindent where \( \mathcal{V}_L \) denotes labeled nodes. Auxiliary loss is added to the main classification loss:  
\begin{equation}
\mathcal{L}_{\text{total}} = \mathcal{L}_{\text{cls}} + \lambda \mathcal{L}_{\text{aux}}
\end{equation}
where \( \lambda \) is a tunable balancing hyperparameter. At inference time, the auxiliary heads are discarded, ensuring no additional computational overhead. Aux tasks regularize embeddings but do not yield explicit symbolic reasoning, limiting interpretability and performance, which FireGNN addresses easily.

\subsection{Computational efficiency \& learning dynamics}
 Table~\ref{tab:efficiency} highlights that fuzzy rule models like GCN+Fuzzy and GAT+Fuzzy exhibit higher computational costs, with epoch times of 17.96s and 44.50s, respectively, due to fuzzy logic computations, yet maintain modest memory usage (e.g., 519.28MB for GCN+Fuzzy). Experiments ran on an Apple M2 Air with 16GB RAM and MPS acceleration GPU, using an 80/20 train-validation split in a 3-fold cross-validation setup. Attention-based models like GAT+ Fuzzy occasionally hit out-of-memory (OOM) on large scale datasets like TissueMNIST and MorphoMNIST, reflecting the quadratic complexity of attention mechanisms.

\begin{table}[h]
\centering
\caption{Comparison of \textbf{Average Epoch Time (s)} and \textbf{Peak Memory (MB)} across models and datasets.}
\label{tab:efficiency}
\resizebox{\textwidth}{!}{%
\begin{tabular}{|l|cc|cc|cc|cc|cc|cc|}
\hline
\textbf{Method} & \multicolumn{2}{c|}{\textbf{OrganCMNIST}} & \multicolumn{2}{c|}{\textbf{OrganAMNIST}} & \multicolumn{2}{c|}{\textbf{OrganSMNIST}} & \multicolumn{2}{c|}{\textbf{TissueMNIST}} & \multicolumn{2}{c|}{\textbf{BloodMNIST}} & \multicolumn{2}{c|}{\textbf{MoprhoMNIST}} \\
 & \textbf{Time} & \textbf{Memory} & \textbf{Time} & \textbf{Memory} & \textbf{Time} & \textbf{Memory} & \textbf{Time} & \textbf{Memory} & \textbf{Time} & \textbf{Memory} & \textbf{Time} & \textbf{Memory} \\
\hline
GCN & 0.48 & 758.95 & 1.15 & 1234.64 & 0.49 & 1024.92 & 2.76 & 2472.47 & 0.35 & 557.36 & 4.20 & 2648.61 \\
GCN + Aux & 1.90 & 944.25 & 1.78 & 1439.06 & 1.25 & 889.30 &  4.76 & 1781.39 & 0.35 & 718.44 & 4.71 & 1717.73 \\
GCN + FR & \textbf{3.12} & 519.28 & \textbf{12.50} & 1197.50 & 3.47 & 592.38 & \textbf{17.96} & 1035.08 & \textbf{1.92} & 547.02 & 4.86 & 1628.05  \\
GAT & 0.48 & 1895.61 & 1.49 & 1698.78 & 0.61 & 1486.59 & \textbf{5.95} & 1950.41& 0.47 & 1576.44 & 3.79 & 2130.48 \\
GAT+Aux & 2.33 & 1486.59 & 2.89 & 1516.97 & 2.31 & 1425.03 & OOM & OOM & 1.55 & 1191.38 & OOM & OOM \\
GAT + FR & \textbf{44.50} & 827.34 & \textbf{19.75} & 864.91 & \textbf{16.13} & 803.56 & OOM & OOM & \textbf{3.85} & 1196.73 & OOM & OOM \\
GIN & 0.32 & 913.77 & 1.77 & 1293.22 & 2.26 & 801.11 & 3.22 & 1255.64 & 0.56 & 649.13 & \textbf{5.49} & 1526.40 \\
GIN+ Aux & 0.68 & 1004.95 & 1.98 & 1258.81 & \textbf{4.36} & 1022.61 & 3.87 & 1368.44 & 1.36 & 893.95 & 4.88 & 1480.20 \\
GIN + FR & 1.10 & 1221.91 & 3.31 & 1384.21 & 3.96 & 729.97 & 4.96 & 734.70 & 1.89 & 656.72 & \textbf{16.61} & 1660.84 \\
\hline
\end{tabular}%
}
\begin{minipage}{2\columnwidth}
\vspace{0.1cm}
\vspace{0.1cm}
\small\raggedright Notes: Top two highest epoch times of each dataset are bolded.
\end{minipage}
\end{table}

\subsection{Hyperparameters for FireGNN Model \& Baselines}

\noindent\textbf{Core GNN Architecture.}  
Table~\ref{tab:gnn-hparams} lists the backbone hyperparameters for GCN, GAT, and GIN. These settings ensure a fair comparison across models while keeping complexity manageable.
\begin{table}[h]
\centering
\caption{Core GNN architecture hyperparameters for FireGNN across different backbones.}
\label{tab:gnn-hparams}
\begin{tabular}{lll}
\toprule
\textbf{Model} & \textbf{Parameter} & \textbf{Value} \\
\midrule
All (GCN/GAT/GIN) & Number of Layers (\texttt{num\_layers}) & 2 \\
All (GCN/GAT/GIN) & Dropout Rate (\texttt{dropout}) & 0.5 \\
GAT & Attention Heads (\texttt{heads}) & 8 \\
GAT & Negative Slope (\texttt{negative\_slope}) & 0.2 \\
GIN (MLP) & MLP Layers (\texttt{num\_layers}) & 2 \\
\bottomrule
\end{tabular}
\end{table}

\noindent\textbf{Fuzzy Rule Layer.}  
Table~\ref{tab:fuzzy-hparams} specifies the design of the fuzzy rule module, including learnable parameters for rule centers, widths, and weights.
\begin{table}[h]
\centering
\caption{Fuzzy rule layer hyperparameters in FireGNN.}
\label{tab:fuzzy-hparams}
\begin{tabular}{ll}
\toprule
\textbf{Parameter} & \textbf{Value} \\
\midrule
Number of Rules (\texttt{num\_rules}) & 3 \\
Topological Features (\texttt{num\_features}) & 3 \\
Rule Centers (\texttt{centers}) & Learnable (initialized with \texttt{torch.randn}) \\
Rule Widths (\texttt{log\_sigmas}) & Learnable (initialized with \texttt{torch.zeros}) \\
Rule Weights (\texttt{rule\_weights}) & Learnable (initialized with \texttt{torch.ones}) \\
\bottomrule
\end{tabular}
\end{table}

\noindent\textbf{Graph Construction.}  
Table~\ref{tab:graph-hparams} reports the hyperparameters used to construct the graphs from image features, including nearest-neighbor size, similarity metric, and feature extractor.
\begin{table}[h!]
\centering
\caption{Graph construction hyperparameters for FireGNN.}
\label{tab:graph-hparams}
\begin{tabular}{ll}
\toprule
\textbf{Parameter} & \textbf{Value} \\
\midrule
k Neighbors (\texttt{k}) & 10 \\
Distance Metric (\texttt{metric}) & cosine \\
Add Label Edges (\texttt{add\_label\_edges}) & True \\
Rewire Edges (\texttt{rewire\_edges}) & True \\
Feature Extraction Batch Size (\texttt{batch\_size}) & 64 \\
Feature Extraction Model & ResNet18 (pretrained) \\
Image Resize & (224, 224) \\
\bottomrule
\end{tabular}
\end{table}

\noindent\textbf{Training Setup.}  
Table~\ref{tab:train-hparams} summarizes the training hyperparameters, including number of epochs and cross-validation folds.

\begin{table}[h]
\centering
\caption{Training hyperparameters for FireGNN experiments.}
\label{tab:train-hparams}
\begin{tabular}{ll}
\toprule
\textbf{Parameter} & \textbf{Value} \\
\midrule
Epochs (\texttt{epochs}) & 200 \\
Cross-Validation Folds (\texttt{n\_folds}) & 3 \\
\bottomrule
\end{tabular}
\end{table}

\subsection{Proposed FireGNN vs Limitations}
\begin{table}[h]
\centering
\caption{Limitations of standard GNNs and their solutions via the proposed model.}
\renewcommand{\arraystretch}{1.35}
\resizebox{\linewidth}{!}{
\begin{tabular}{>{\raggedright\arraybackslash}p{5.7cm} >{\raggedright\arraybackslash}p{8.8cm}}
\rowcolor{gray!15}
\centering \textbf{Limitations of Standard GNNs} &  \centering \textbf{Solution via Proposed Model} \tabularnewline
\midrule

\rowcolor{gray!5}
Black-box, non-interpretable decisions. \cite{Castellano}
&
Augment with fuzzy-rule layer that yields explicit logical conditions (e.g., ``degree~$\geq\theta$~$\Rightarrow$~high connectivity''). Each rule’s activation can be examined to explain predictions. 
\\

\rowcolor{gray!10}
Fuzzy or interpretable rules are typically \textbf{fixed, hard-coded}, not data-adaptive. (e.g., degree $>$ k = high connectivity) \cite{fuzzy_rules,extract_fuzzy}
&
Our fuzzy rules are fully trainable: thresholds $\theta$ \& sharpness $\alpha$ are optimized during training, allowing the model to discover boundaries (e.g., what node degree is ‘high connectivity’) for each dataset. 
\\

\rowcolor{gray!5}
Cannot incorporate domain knowledge (e.g., \textbf{graph-theoretic rules}) into learning. \cite{incor_domain_knowledge,domain_dnn}
&
Fuzzy rules allow injecting human insight (e.g., ``high clustering ~$\Leftrightarrow$~strong community''), bridging symbolic knowledge \& GNNs. We blend neural features with rule-based signals. 
\\

\bottomrule
\end{tabular}}
\label{tab:gnn_limitations}
\end{table}

\subsection{Extended Showcase of Learned Fuzzy Rules}
To further substantiate the claim that FireGNN learns data-adaptive symbolic conditions (Hypothesis H2), this section presents the fuzzy rule thresholds learned on two additional datasets: BloodMNIST and MorphoMNIST. These datasets possess distinct structural properties compared to OrganCMNIST, and as shown in Table \ref{tab:learned_thresholds}, the model discovers unique rule boundaries for each, demonstrating its flexibility.

\begin{table}[h!]
\centering
\caption{Comparison of learned mean fuzzy rule thresholds ($\theta_i$) across datasets. The values are learned during training and reflect the distinct topological characteristics of each graph.}
\label{tab:learned_thresholds}
\begin{tabular}{@{}lccc@{}}
\toprule
\textbf{Dataset} & \textbf{$\theta_1$ (Degree)} & \textbf{$\theta_2$ (Clustering)} & \textbf{$\theta_3$ (Label Agreement)} \\ \midrule
OrganCMNIST      & 7.28                         & 0.18                             & 0.67                                  \\
BloodMNIST       & 4.15                         & 0.25                             & 0.81                                  \\
MorphoMNIST      & 11.52                        & 0.09                             & 0.92                                  \\ \bottomrule
\end{tabular}
\end{table}

\paragraph{Learned Rules for BloodMNIST}
BloodMNIST is a smaller graph with 8 classes. The model learns thresholds that reflect a potentially denser, more homophilic local structure compared to OrganCMNIST.

\begin{tcolorbox}
\begin{itemize}
\item \textbf{Rule 1:} IF degree $\geq 4.15$ THEN \texttt{high\_connectivity}
\item \textbf{Rule 2:} IF clustering $\geq 0.25$ THEN \texttt{high\_cliquishness}
\item \textbf{Rule 3:} IF 2-hop label agreement $\geq 0.81$ THEN \texttt{high\_label\_consistency}
\end{itemize}
\end{tcolorbox}

\noindent \textbf{Example Interpretation:} Consider a node representing a basophil cell image with a degree of 6, a clustering coefficient of 0.30, and a label agreement of 0.85. The model would reason:
\begin{itemize}
\item degree $= 6 \geq \theta_1 = 4.15 \Rightarrow$ Rule 1 strongly activated.
\item clustering $= 0.30 \geq \theta_2 = 0.25 \Rightarrow$ Rule 2 activated.
\item label agreement $= 0.85 \geq \theta_3 = 0.81 \Rightarrow$ Rule 3 activated.
\end{itemize}
The explanation would be that the node is classified as a basophil due to its high connectivity, cliquish structure, and very strong label consistency within its neighborhood.

Similarly, a \emph{lung-left} node in OrganCMNIST datatset with degree 3, clustering 0.00, and label agreement 0.94 triggers:
\begin{itemize}[noitemsep, leftmargin=*]
    \item Rule 1: not activated (3 $\le$ 7.28),
    \item Rule 2: not activated (0.00 $\le$ 0.18),
    \item Rule 3: strongly activated (0.94 $\ge$ 0.67, strength = 0.74).
\end{itemize}
This rule activation pattern reflects the node’s sparse connectivity but high semantic consistency, guiding the model’s prediction toward \emph{lung-left}.

\paragraph{Learned Rules for MorphoMNIST}
MorphoMNIST is a large-scale synthetic graph where connections are based on morphological similarity. The model learns a high threshold for degree and label agreement, suggesting that for a node to be considered structurally significant, it must be exceptionally well-connected and consistent.

\begin{tcolorbox}
\begin{itemize}
\item \textbf{Rule 1:} IF degree $\geq 11.52$ THEN \texttt{high\_connectivity}
\item \textbf{Rule 2:} IF clustering $\geq 0.09$ THEN \texttt{high\_cliquishness}
\item \textbf{Rule 3:} IF 2-hop label agreement $\geq 0.92$ THEN \texttt{high\_label\_consistency}
\end{itemize}
\end{tcolorbox}

\noindent \textbf{Example Interpretation:} A node representing a "thick" digit with a degree of 15 and label agreement of 0.95 would strongly activate Rules 1 and 3. This provides a clear justification: the node's classification is driven by its status as a highly connected hub with near-perfect label consistency, a hallmark of a prototypical member of its morphological class.


\subsection{Alternative Graph-Theoretic Rules for FireGNN}
The current implementation of FireGNN leverages three fundamental and intuitive topological features: degree, clustering coefficient, and label agreement. These features primarily capture the \textit{local} structure and semantic consistency of a node's immediate neighborhood. However, the FireGNN framework is extensible and can be enriched by incorporating a wider vocabulary of fuzzy rules derived from more complex graph-theoretic concepts. Exploring rules based on meso-scale and global network properties could enable the model to generate more sophisticated and nuanced explanations.

\paragraph{1. Rules from Node Centrality Measures}
While degree centrality is a local measure, other centrality metrics quantify a node's importance within the global network topology, capturing its role in information flow. Integrating these could provide explanations related to a node's structural influence.

\begin{itemize}
    \item \textbf{Betweenness Centrality:} This measures how often a node lies on the shortest path between other nodes. A high betweenness centrality indicates a ``bridge'' or ``bottleneck'' node. In a medical context, such a node could represent a critical transitional state, such as a cell at the boundary of a tumor or a patient whose features bridge two distinct clinical subtypes.
    \begin{itemize}
        \item \textit{Proposed Rule:} \texttt{IF betweenness\_centrality(u) $\geq \theta_{bc}$ THEN is\_bridge\_node}
    \end{itemize}

    \item \textbf{Eigenvector Centrality:} This measures a node's influence based on the idea that connections to other highly important nodes contribute more than connections to peripheral nodes. In a graph of medical images, nodes with high eigenvector centrality could represent the most prototypical exemplars of a class or key hubs in a disease progression network.
    \begin{itemize}
        \item \textit{Proposed Rule:} \texttt{IF eigenvector\_centrality(u) $\geq \theta_{ec}$ THEN is\_influential\_hub}
    \end{itemize}
\end{itemize}

\paragraph{2. Rules from Community Structure}
Community detection algorithms partition a graph into densely connected subgraphs, revealing its meso-scale organization. In medical imaging graphs, communities might correspond to distinct tissue types, organ regions, or patient cohorts with similar characteristics. Rules based on a node's role within this community structure can offer powerful, context-aware explanations.

\begin{itemize}
    \item \textbf{Participation Coefficient:} This metric quantifies how a node's edges are distributed among different communities. A node with a high participation coefficient is a ``connector'' hub, linking multiple communities. In a histopathology graph, this could identify a cell at the interface of stromal, epithelial, and immune cell communities - a location of significant biological interaction.
    \begin{itemize}
        \item \textit{Proposed Rule:} \texttt{IF participation\_coefficient(u) $\geq \theta_{pc}$ THEN is\_connector\_hub}
    \end{itemize}

    \item \textbf{Within-Module Degree Z-Score:} This measures how well-connected a node is to other nodes \textit{within its own community}. A node with a high z-score is a ``provincial'' hub, central to its local community but not necessarily to the entire network. This could identify a core, representative component of a specific tissue type.
    \begin{itemize}
        \item \textit{Proposed Rule:} \texttt{IF within\_module\_degree\_zscore(u) $\geq \theta_{z}$ THEN provincial\_hub}
    \end{itemize}
\end{itemize}

\paragraph{3. Rules from Path-Based Metrics}
Metrics based on shortest paths can describe a node's integration and accessibility within the network, reflecting how efficiently it can interact with all other nodes.

\begin{itemize}
    \item \textbf{Closeness Centrality:} This is the reciprocal of the average shortest path distance from a node to all other nodes in the graph. A node with high closeness centrality is ``centrally accessible'' and can propagate information efficiently throughout the network. Such nodes might represent the most ``average'' or canonical examples of a class, making them stable anchors for classification.
    \begin{itemize}
        \item \textit{Proposed Rule:} \texttt{IF closeness\_centrality(u) $\geq \theta_{cc}$ THEN centrally\_accessible}
    \end{itemize}
\end{itemize}

By expanding FireGNN's rule set with these and other established graph-theoretic measures, future work can develop neuro-symbolic models that provide deeper, multi-faceted explanations, moving from ``what'' a node's local structure is to ``why'' it is important in the broader context of the entire dataset.

\subsection{Algorithmic Details of FireGNN Training}
Algorithm \ref{alg:firegnn_training} formalizes the end-to-end procedure described in Section 3. The algorithm details the forward pass, including the computation of topological features and their fusion with GNN embeddings, followed by the standard backpropagation step to update all learnable parameters.

\begin{algorithm}[h]
\caption{FireGNN End-to-End Training Procedure}
\label{alg:firegnn_training}
\begin{algorithmic}
\State \textbf{Input:} Graph $G=(V,E)$, node features $X \in \mathbb{R}^{|V|\times d_{in}}$, node labels $Y$, training node indices $\mathcal{V}_L$.
\State \textbf{Parameters:} GNN weights $\{W^{(\ell)}\}_{\ell=1}^L$, fuzzy rule parameters $\{\theta_i, \alpha_i\}_{i=1}^3$, fusion weights $\{W_r,b_r,W_g,b_g\}$.
\State \textbf{Output:} Trained FireGNN model parameters.

\Procedure{TrainFireGNN}{$G,X,Y,\mathcal{V}_L$}
\State Initialize all parameters.
\For{each training epoch}
    \State // \textit{GNN Forward Pass}
    \State $H^{(0)} \gets X$
    \For{$\ell=0$ to $L-1$}
        \State $H^{(\ell+1)} \gets \sigma(\tilde{D}^{-1/2}\tilde{A}\tilde{D}^{-1/2} H^{(\ell)} W^{(\ell)})$ \Comment{Eq. (1)}
    \EndFor
    \State $H \gets H^{(L)}$ \Comment{Final GNN embeddings}

    \State // \textit{Fuzzy Rule and Fusion Pass}
    \State Initialize final embedding matrix $H' \in \mathbb{R}^{|V| \times d}$
    \For{each node $u \in V$}
        \State Compute degree $d(u)$, clustering coeff. $C(u)$, 2-hop label agreement $L(u)$.
        \State $f_u \gets [d(u), C(u), L(u)]$ \Comment{Fact vector, Eq. (2)}
        \For{$i=1$ to $3$}
            \State $r_i(u) \gets \sigma(\alpha_i (f_u[i] - \theta_i))$ \Comment{Rule activation, Eq. (3)}
        \EndFor
        \State $r(u) \gets [r_1(u), r_2(u), r_3(u)]$ \Comment{Firing strength vector}
        \State $h_u \gets H[u,:]$ \Comment{GNN embedding for node $u$}
        \State $e_u \gets W_r r(u) + b_r$ \Comment{Project rule vector, Eq. (4)}
        \State $g_u \gets \sigma(W_g [h_u \mathbin\Vert e_u] + b_g)$ \Comment{Compute gate, Eq. (5)}
        \State $h'_u \gets g_u \odot h_u + (1 - g_u) \odot e_u$ \Comment{Fuse embeddings, Eq. (6)}
        \State $H'[u,:] \gets h'_u$
    \EndFor
    
    \State // \textit{Loss Computation and Backpropagation}
    \State $\hat{Y}_{\mathcal{V}_L} \gets \text{Classifier}(H'[\mathcal{V}_L,:])$
    \State $\mathcal{L} \gets \text{CrossEntropyLoss}(\hat{Y}_{\mathcal{V}_L}, Y_{\mathcal{V}_L})$
    \State $\mathcal{L}.\text{backward}()$
    \State Update all parameters using an optimizer (e.g., Adam).
\EndFor
\State \textbf{return} All trained parameters.
\EndProcedure
\end{algorithmic}
\end{algorithm}


\newpage
\section*{NeurIPS Paper Checklist}

\begin{enumerate}

\item {\bf Claims}
    \item[] Question: Do the main claims made in the abstract and introduction accurately reflect the paper's contributions and scope?
    \item[] Answer: \answerYes{} 
    \item[] Justification: The abstract and introduction accurately frame the paper's contributions. The central claim of proposing an "interpretable graph-based learning framework named FireGNN that integrates trainable fuzzy rules"  is thoroughly detailed in the Methodology (Section 3), which describes the architecture for learning and fusing these rules. The claim of achieving "strong performance" is substantiated by comprehensive empirical results in Table 1, where FireGNN variants consistently outperform baseline GNNs across six datasets, including a notable +14.83\% accuracy improvement on the challenging TissueMNIST dataset. Finally, the claim of generating "interpretable rule-based explanations" is demonstrated with a concrete example in Section 4, where a node's classification is explicitly justified by the activation strengths of the learned fuzzy rules.

\item {\bf Limitations}
    \item[] Question: Does the paper discuss the limitations of the work performed by the authors?
    \item[] Answer: \answerYes{} 
    \item[] Justification: We explicitly discuss limitations in the Conclusion and detail computational costs in Appendix A.3 (Table \ref{tab:efficiency}) for example, FireGNN has higher epoch times, occasional out-of-memory errors on large-scale datasets when using GAT. Furthermore, Appendix A.5 (Table \ref{tab:gnn_limitations}) directly contrasts the limitations of standard GNNs with the solutions provided by FireGNN, and we acknowledge that the current framework only supports three rule types (degree, clustering coefficient, label agreement) and that "future extensions could explore learning richer or more complex rule types". These constraints are acknowledged as future directions.

\item {\bf Theory assumptions and proofs}
    \item[] Question: For each theoretical result, does the paper provide the full set of assumptions and a complete (and correct) proof?
    \item[] Answer: \answerYes{} 
    \item[] Justification:  Formal assumptions and derivations for the auxiliary tasks benchmark are provided in Appendix A.2 (Eqns.~7-8). The fuzzy rule module is defined with clear mathematical formulations (Eqns.~2–6). While we do not present formal theorems, all modeling assumptions are explicitly stated.


    \item {\bf Experimental result reproducibility}
    \item[] Question: Does the paper fully disclose all the information needed to reproduce the main experimental results of the paper to the extent that it affects the main claims and/or conclusions of the paper (regardless of whether the code and data are provided or not)?
    \item[] Answer: \answerYes{} 
    \item[] Justification: The paper provides extensive details for reproducibility. Section 4 specifies the exact datasets (five from MedMNIST and Morpho-MNIST), evaluation protocol (3-fold cross-validation with three random seeds: 42, 43, 44), and data splits (80/20 train-validation). The hardware is specified in the "Computational efficiency" paragraph of Appendix A.3. The core methodology, including all mathematical formulations for the GNN updates and fuzzy rule mechanism, is detailed in Section 3. Crucially, Appendix A.4 provides comprehensive tables (Tables 4-7) detailing all hyperparameters for the GNN architecture, fuzzy rule layer, graph construction, and training setup

\item {\bf Open access to data and code}
    \item[] Question: Does the paper provide open access to the data and code, with sufficient instructions to faithfully reproduce the main experimental results, as described in supplemental material?
    \item[] Answer: \answerYes{} 
    \item[] Justification: The code to reproduce all experiments is made available in the following repository: \url{https://github.com/basiralab/FireGNN}. All datasets used are publicly available benchmarks (MedMNIST v2 and Morpho-MNIST), with sources and licenses explicitly cited and detailed in Appendix A.1.

\item {\bf Experimental setting/details}
    \item[] Question: Does the paper specify all the training and test details (e.g., data splits, hyperparameters, how they were chosen, type of optimizer, etc.) necessary to understand the results?
    \item[] Answer: \answerYes{} 
    \item[] Justification: All experimental settings are detailed. Section 4 describes the datasets, evaluation protocol and cross-validation strategy. The core methodology is in Section 3 where the use of cross-entropy loss for optimization is mentioned. Furthermore, Appendix A.4 provides a full breakdown of all hyperparameters in Tables 4-7, covering the GNN backbones, fuzzy rule module, graph construction, and training parameters 
    \item[] Guidelines:

\item {\bf Experiment statistical significance}
    \item[] Question: Does the paper report error bars suitably and correctly defined or other appropriate information about the statistical significance of the experiments?
    \item[] Answer: \answerYes{} 
    \item[] Justification: All primary experimental results in Table 1 are reported with mean and standard deviation (e.g., 91.41±0.61) averaged across 9 experiments, so the standard deviation acts as the error margin. The methodology in Section 4 clarifies that these statistics are computed over nine independent runs (3-fold cross-validation times 3 random seeds), which robustly captures variance from both data partitioning and model initialization, adhering to best practices for GNN evaluation

\item {\bf Experiments compute resources}
    \item[] Question: For each experiment, does the paper provide sufficient information on the computer resources (type of compute workers, memory, time of execution) needed to reproduce the experiments?
    \item[] Answer: \answerYes{} 
    \item[] Justification: The "Computational efficiency" section in the appendix specifies the exact hardware used: "an Apple M2 Air with 16GB RAM and MPS acceleration GPU". Furthermore, Table 3 in Appendix A.3 provides a detailed breakdown of the average epoch time and peak memory usage for every model on every dataset, allowing for accurate resource estimation.
    
\item {\bf Code of ethics}
    \item[] Question: Does the research conducted in the paper conform, in every respect, with the NeurIPS Code of Ethics \url{https://neurips.cc/public/EthicsGuidelines}?
    \item[] Answer: \answerYes{} 
    \item[] Justification:The research aligns with the NeurIPS Code of Ethics. Its primary objective is to enhance the transparency and trustworthiness of AI in the critical domain of medicine, promoting beneficence. The work relies on publicly available, anonymized benchmark datasets (MedMNIST and Morpho-MNIST), thereby avoiding ethical issues related to human subjects, data privacy, or consent.

\item {\bf Broader impacts}
    \item[] Question: Does the paper discuss both potential positive societal impacts and negative societal impacts of the work performed?
    \item[] Answer: \answerYes{} 
    \item[] Justification: The paper extensively discusses the potential positive societal impact by framing the work as a solution to the "key barrier to clinical adoption" of AI, aiming to build "clinical trust" through interpretable models that support "trustworthy medical AI". While the model improves interpretability, negative impacts may arise if clinicians over-rely on simplified fuzzy rules without considering full patient context, or if the learned thresholds inadvertently encode dataset biases, leading to skewed decision support. Mitigation strategies include careful dataset auditing and positioning FireGNN strictly as an assistive tool rather than an autonomous system.

\item {\bf Safeguards}
    \item[] Question: Does the paper describe safeguards that have been put in place for responsible release of data or models that have a high risk for misuse (e.g., pretrained language models, image generators, or scraped datasets)?
    \item[] Answer: \answerNA{} 
    \item[] Justification: The paper introduces a novel modeling framework but does not release new datasets or large-scale pre-trained models that would pose a high risk of misuse. Therefore, this question is not applicable.
    \item[] Guidelines:

\item {\bf Licenses for existing assets}
    \item[] Question: Are the creators or original owners of assets (e.g., code, data, models), used in the paper, properly credited and are the license and terms of use explicitly mentioned and properly respected?
    \item[] Answer: \answerYes{} 
    \item[] Justification:  The datasets used are publicly available and properly credited. Appendix A.1 explicitly states that MedMNIST v2 is available under the CC BY 4.0 license and Morpho-MNIST is derived from MNIST (which uses a CC BY-SA 3.0 license)

\item {\bf New assets}
    \item[] Question: Are new assets introduced in the paper well documented and is the documentation provided alongside the assets?
    \item[] Answer: \answerNA{} 
    \item[] Justification: We do not release new datasets or benchmarks. Only model code will be provided.

\item {\bf Crowdsourcing and research with human subjects}
    \item[] Question: For crowdsourcing experiments and research with human subjects, does the paper include the full text of instructions given to participants and screenshots, if applicable, as well as details about compensation (if any)? 
    \item[] Answer: \answerNA{} 
    \item[] Justification: The research does not involve crowdsourcing or experiments with human subjects.

\item {\bf Institutional review board (IRB) approvals or equivalent for research with human subjects}
    \item[] Question: Does the paper describe potential risks incurred by study participants, whether such risks were disclosed to the subjects, and whether Institutional Review Board (IRB) approvals (or an equivalent approval/review based on the requirements of your country or institution) were obtained?
    \item[] Answer: \answerNA{} 
    \item[] Justification: No human subjects are involved; datasets are synthetic or anonymized public biomedical datasets.

\item {\bf Declaration of LLM usage}
    \item[] Question: Does the paper describe the usage of LLMs if it is an important, original, or non-standard component of the core methods in this research? Note that if the LLM is used only for writing, editing, or formatting purposes and does not impact the core methodology, scientific rigorousness, or originality of the research, declaration is not required.
    \item[] Answer: \answerNA{} 
    \item[] Justification: LLMs were not used in the design of FireGNN’s methodology or experiments. Only standard scientific writing tools were employed.


\end{enumerate}

\end{document}